\documentclass[reprint,aps,prl,amsmath,amssymb,twocolumn,superscriptaddress]{revtex4-2}

\usepackage[utf8]{inputenc}
\usepackage{graphicx}%
\graphicspath{ {./Figures/} }
\usepackage{multirow}%
\usepackage{amsmath,amssymb,amsfonts}%
\usepackage{mathrsfs}%
\usepackage{mathtools}
\usepackage[title]{appendix}%
\usepackage{xcolor}%
\usepackage{orcidlink}
\usepackage{CJKutf8}
\usepackage[capitalise]{cleveref}

\newcommand{\PRLSec}[1]{\emph{#1.---}}
\begin{document}
\raggedbottom

\title{Beyond geometric symmetry: Broadband linear relations in wave scattering}
\begin{CJK}{UTF8}{gbsn}
\author{Malte Röntgen \orcidlink{0000-0001-7784-8104}}
\email{mroentgen@eitech.edu.cn}
\affiliation{Eastern Institute for Advanced Study, Eastern Institute of Technology, Ningbo, China}

\author{Sucui Luo (罗素翠)\orcidlink{0009-0007-1766-4558}}
\email{sucuiluo92@gmail.com}
\affiliation{MOE Key Laboratory of Advanced Micro-Structured Materials, Shanghai Frontiers Science Center of Digital Optics, Institute of Precision Optical Engineering, and School of Physics Science and Engineering, Tongji University, Shanghai 200092, China}

\author{Xuelong Chen (陈学龙) \orcidlink{0009-0002-3885-5402}}
\affiliation{Eastern Institute for Advanced Study, Eastern Institute of Technology, Ningbo, China}
\affiliation{Department of Electrical and Electronic Engineering, The Hong Kong Polytechnic University, Hong Kong, China}

\author{Chenyu Zhang (张晨宇)\orcidlink{0009-0005-1469-4511}}
\affiliation{MOE Key Laboratory of Advanced Micro-Structured Materials, Shanghai Frontiers Science Center of Digital Optics, Institute of Precision Optical Engineering, and School of Physics Science and Engineering, Tongji University, Shanghai 200092, China}

\author{Ruotao Ye (叶若涛) \orcidlink{0009-0000-0044-219X}}
\affiliation{Eastern Institute for Advanced Study, Eastern Institute of Technology, Ningbo, China}
\affiliation{School of Physics and Astronomy, Shanghai Jiao Tong University, Shanghai, China}

\author{Tianshu Jiang (姜天舒)\orcidlink{0000-0002-0157-3877}}
\email{tsjiang@tongji.edu.cn}
\affiliation{MOE Key Laboratory of Advanced Micro-Structured Materials, Shanghai Frontiers Science Center of Digital Optics, Institute of Precision Optical Engineering, and School of Physics Science and Engineering, Tongji University, Shanghai 200092, China}

\author{Wenlong Gao (高文龙) \orcidlink{0000-0002-3446-7157}}
\email{wgao@eitech.edu.cn}
\affiliation{Eastern Institute for Advanced Study, Eastern Institute of Technology, Ningbo, China}

\begin{abstract}
The design and control of wave scattering, that is, of the reflection and transmission parameters of a device, is of ubiquitous importance.
These parameters generally change with varying frequency, though certain \emph{frequency-independent} linear relations may exist between them.
Reciprocity and geometric symmetry (reflections, rotations, etc.) are classic and well-known examples that are present in many devices and significantly ease their design.
In this work, we go beyond these and introduce a new class of relations that cannot be induced by reciprocity or geometric symmetry.
Choosing networks of waveguides as our workhorse, we discuss the conditions and consequences of such novel behaviour and showcase suitable example setups.
We further experimentally test our predictions using coaxial cables and find excellent agreement in the broad frequency range between 0 and 1 GHz.
Our work not only deepens the theoretical understanding of waveguide network dynamics, but also opens new avenues for applications in broadband signal processing and integrated photonics.
\end{abstract}

\maketitle
\end{CJK}

\PRLSec{Introduction}%
The ability to control the scattering of waves lies at the heart of a wide range of physical systems \cite{newtonScatteringTheoryWaves1982}, from microwave circuits \cite{Pozar2012MicrowaveEngineering} and integrated photonic devices to acoustic metamaterials \cite{Jimenez2021143AcousticWavesPeriodicStructures} and quantum systems \cite{Datta1995ElectronicTransportMesoscopicSystems}.
In these settings, the reflection and transmission amplitudes of a scatterer determine how information, energy, or particles propagate through a device and therefore constitute fundamental design parameters. While these scattering properties generally depend on frequency, they are often constrained by frequency-independent relations that originate from fundamental physical principles. Reciprocity, conservation laws leading to unitarity constraints, and geometric symmetries are prominent examples, providing powerful restrictions on the scattering matrix that both simplify theoretical analyses and guide the design of functional devices. Identifying new classes of such constraints is of fundamental interest, as they can reveal previously unexplored structures in wave dynamics and enable functionalities beyond those attainable from conventional symmetry principles.

Generally speaking, for a device with $N$ ports and geometric symmetry---such as a reflection---, any ports that are mapped onto one another by that symmetry are indistinguishable; their reflection coefficients are identical for any frequency.
Recently, it was shown that such broadband equireflectionality can occur even in asymmetric setups, provided that there is a \emph{hidden} geometric symmetry \cite{Rontgen2023PRA20044082EquireflectionalityCustomizedUnbalancedCoherent,Sol2024AM362303891CovertScatteringControlMetamaterials}.
That is, the seemingly asymmetric system becomes geometrically symmetric after performing a dimensional reduction akin to an effective Hamiltonian \cite{Smith2019PA514855HiddenSymmetriesRealTheoretical,Bunimovich2014IsospectralTransformationsNewApproach}.

Motivated by these results, we wondered whether one can achieve something that cannot be achieved by a geometric symmetry.
That is, a broadband, linear relation between the scattering parameters that involves both the reflection ($r_i$) and transmission ($t_{ij}$)  coefficients.  
In the simplest case of a reciprocal two-port network, this relation reads $r_2(f) = r_1(f) + \alpha\, t_{12}(f)$, with some frequency-independent scalar $\alpha \ne 0$.

We will start this work by presenting a surprisingly simple example setup for this phenomenon.
As we show, it features a non-geometric symmetry, which puts strong constraints on all eigenmodes and, as a consequence, leads to the above linear relation between the scattering parameters.
From there, we generalize our results to networks with a more complicated structure, again leading to linear relations between scattering parameters that hold for all frequencies and for more than two ports.
Finally, we experimentally test our predictions for a two-port setup using coaxial cables and find excellent agreement in the broad frequency range between 0 and 1 GHz.

We emphasize that, though our results are presented in terms of waveguide networks, the underlying logic is universal; it is to tailor a setup such that all of its eigenmodes fulfil some non-trivial relation on a (small) set of (point-like) domains, and then using this relation by coupling the domains to the environment.

\begin{figure}[!htb]
\includegraphics[width=\linewidth]{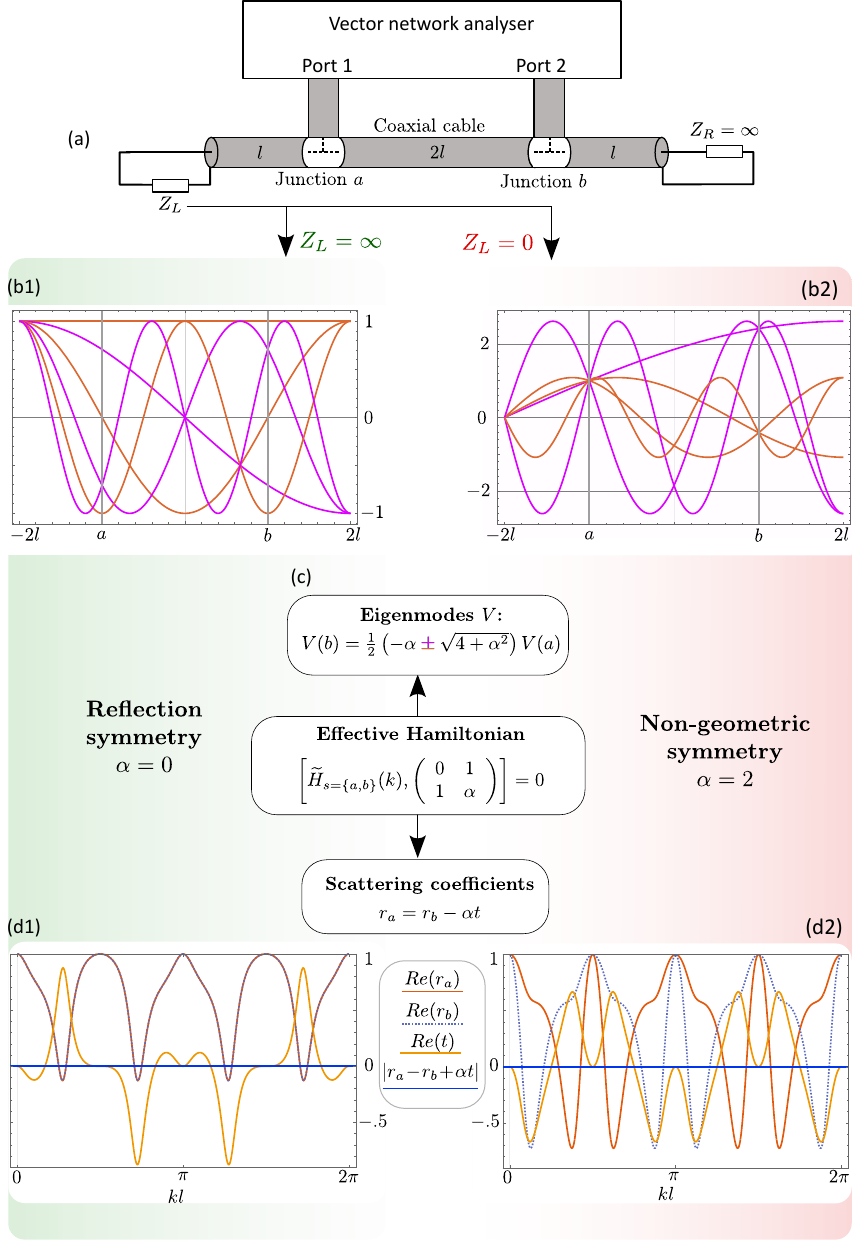}
\caption{\label{fig:Fig1}
(\textbf{a}) Simple resonator, formed by three coaxial cables.
Left panel: Reflection symmetric choice of $Z_L = Z_R$, right panel: Choosing $Z_L = 0$ breaks geometric symmetry, but induces a non-geometric symmetry (see text for details). The (non-)geometric symmetry induces relations on both eigenmodes and scattering parameters (\textbf{c}), depicted graphically in (\textbf{b1}), (\textbf{b2}), and (\textbf{d1}), (\textbf{d2}), respectively.
In (\textbf{b1}), (\textbf{b2}), lines colours red and purple are used to mark modes with positive and negative (point-wise scaled) parity, respectively.
} 
\end{figure} 
\PRLSec{Introductory example}
Broadband relations between scattering parameters can occur in surprisingly simple setups, as we now demonstrate.
To begin, consider the setup shown in \cref{fig:Fig1}(a), consisting of three coaxial cables of respective lengths $l$, $2l$, and $l$, joined by connectors.
Without loss of generality, we set $l=1$.

Let us first analyse the closed system.
Assuming ideal connectors, the setup is nothing but a resonator formed by a single transmission line of length $4$.
To find its eigenmodes, we switch to the frequency domain; the voltage wave field $V(x)$ then fulfils
\begin{equation} \label{eq:1DHelmholtz}
    \frac{d^{2}V(x)}{dx^{2}} + k^2 V(x) = 0 \,
\end{equation}
with $k = 2\pi f/v_g$, with $f$ the frequency and $v_g$ the speed of light inside the cable. 
The setup's behaviour crucially depends on how we terminate the left and right cable ends.

To set the stage, we first investigate the trivial case of leaving both cable ends open.
This corresponds to terminating them with $Z_{L,R} = \infty$, such that the current $I \propto V'$ vanishes at both ends.
With the setup being symmetric, all eigenmodes have even or odd parity; this can also be seen from \cref{fig:Fig1}(b1), where we show the first six eigenmodes.
To investigate the scattering characteristics, we replace the two-port junctions at $x = -1:=a$, $x = 1 := b$ by $T$-junctions, connect these to a vector network analyser and measure the device's scattering matrix $S(k)$.
Since the setup is reciprocal, we have $ S(k) = S^T(k)=  \begin{psmallmatrix}
        r_a & t \\
        t& r_b
        \end{psmallmatrix}$.
Additionally, due to mirror-symmetry of the setup, we have $r_a = r_b:=r$ for all frequencies [\cref{fig:Fig1}(d1)].
While powerful, this relation is quite limited in scope, as it only involves the reflection coefficients $r_1$ and $r_2$.
In the following, we will show that by breaking the mirror symmetry, we can obtain a linear relation that involves both the reflection and transmission coefficients, $r_1$, $r_2$, and $t$.

We break the symmetry by short-circuiting the left end of the cable, which corresponds to terminating it with $Z_L = 0$, such that the voltage $V$ vanishes there.
In \cref{fig:Fig1}(b2), we show the first six eigenmodes of the modified resonator; they no longer have definite parity.
Crucially, however, they have the rather unusual property of \emph{point-wise scaled parity} at the special locations $a,b$,
\begin{equation} \label{eq:scaledParity}
    V(b) = \left(1 \pm \sqrt{2} \right) \, V(a) \,,
\end{equation}
which, as we shall show in the following, leads to the broadband linear relation
\begin{equation} \label{eq:scatteringRelation}
    r_a = r_b - 2 t
\end{equation}
between the scattering parameters \emph{for all frequencies}; cf. \cref{fig:Fig1}(d2).

\PRLSec{Broadband non-geometric symmetries}%
To understand the above findings, we start with \cref{eq:scaledParity}, that is, we analyse the closed system.
Generally speaking, a network of single-mode waveguides (such as coaxial cables) can be described by a non-linear eigenvalue problem 
\begin{equation} \label{eq:nonlinearEVP}
    \mathcal{H}_s(k) \vec{V}_s = 0
\end{equation}
with $k$ the eigenfrequency,
\begin{equation*}
    \vec{V}_s = \left( V(s_1), \ldots, V(s_n) \right)^T
\end{equation*}
the $n$-dimensional vector obtained by sampling the full eigenmode $V$ at locations $s = s_1,\ldots,s_n$, and $\mathcal{H}_s(k)$ the effective Hamiltonian that describes the dynamics on $s$; see End Matter for details on deriving $\mathcal{H}_s$.
We note that the non-linearity of \cref{eq:nonlinearEVP} is with respect to $k$; it is to be distinguished from non-linearities in the eigenvector (wave field) itself, which occurs, for instance, in non-linear photonics.

For a given network, many different effective Hamiltonians exist, as there are infinitely many choices for the sampling points $s$.
Luckily, this problem of choice is irrelevant here; since we are interested in understanding \cref{eq:scaledParity}, a natural choice is $s = \{a,b\}$.
Some algebra then gives us
\begin{equation} \label{eq:simpleStructure}
    \mathcal{H}_s(k) =  \beta(k) I +  \gamma_2(k) B
\end{equation}
with
$\beta = -\cot(k) - \cot(2 k)$,
$\gamma_n = 1/\sin(n k)$, $I$ the identity matrix, and $B = 
\begin{psmallmatrix}
 0 & 1 \\
 1 & 2 \\
\end{psmallmatrix}$.

We say that an effective Hamiltonian features a \emph{broadband symmetry $Q$} if
\begin{equation} \label{eq:effectiveCommutation}
    \mathcal{H}_s(k) Q = Q \mathcal{H}_s(k) \quad \forall \, k \,.
\end{equation}
Such a symmetry has a strong impact: In the absence of degeneracies, the eigenvectors of $\mathcal{H}_s$---and thus the eigenmodes $V$ of the full network, sampled at locations $s$---are equal to the eigenvectors of $Q$ and, as a consequence, \emph{frequency-independent}.
For our system of \cref{fig:Fig1} with $Z_L = 0$, we have $Q = B$.
The eigenvectors of that matrix are $(1, 1 \pm \sqrt{2})^T$; this explains \cref{eq:scaledParity}.

Having understood the behaviour of eigenmodes, we continue with scattering.
When opening the system at points $s$, a broadband symmetry of $\mathcal{H}_s$ translates into a broadband symmetry of the corresponding scattering matrix $S_s$.
To see this, note that the latter can be computed as
\cite{Hofmann2021PRE104045211SpectralDualityGraphsMicrowave}
\begin{equation} \label{eq:sMat}
    S_s(k) = (I + i \mathcal{H}_s^{-1})^{-1} (I - i \mathcal{H}_s^{-1}) \,.
\end{equation}
After some algebra, it follows that
\begin{equation} \label{eq:commutator}
    \left[\mathcal{H}_s, Q \right] =0 \Rightarrow \left[S_s, Q \right] = 0
\end{equation}
where $[X,Y] = XY - YX$ denotes the commutator, and where both equations hold for all frequencies $k$.

Multiplying out \cref{eq:commutator}, we see that a broadband symmetry imposes one or more linear relations on the scattering parameters.
For our case of $Q = B$ and two ports, $S_s(k) = \begin{psmallmatrix}
    r_a & t \\ t & r_b
\end{psmallmatrix}$, we obtain the linear relation $r_a = r_b - 2t$; this is exactly \cref{eq:scatteringRelation}.

To summarize, we have traced the origin of scaled point-wise parity, \cref{eq:scaledParity}, as well as the broadband relation between scattering parameters, to the broadband symmetry $Q = B$.
To understand this symmetry better, it is instructive to repeat the above steps for the mirror-symmetric setup where $Z_L = Z_R = \infty$.
Sampling eigenmodes at the two points $a,b$, we derive an effective Hamiltonian 
\begin{equation}
 \mathcal{H}_s = I \xi(k) + \gamma_1(k) R
\end{equation}
with $\xi = \beta(k) + \gamma_1 / \cos(k)$, and with the $2\times{2}$ reflection matrix $R = 
\begin{psmallmatrix}
 0 & 1 \\
 1 & 0 \\
\end{psmallmatrix}$.
It follows that this effective Hamiltonian fulfils \cref{eq:effectiveCommutation,eq:commutator} for $Q:=R$, and we see that $r_a = r_b$, as expected.
For the case of $Z_L = 0$, on the other hand, the effective Hamiltonian commutes with $\begin{psmallmatrix}
 0 & 1 \\
 1 & 2 \\
\end{psmallmatrix}$, which is not a reflection operation.   
Thus, the corresponding effective Hamiltonian features a \emph{non-geometric symmetry}.

\begin{figure}[!htb]
\includegraphics[width=\linewidth]{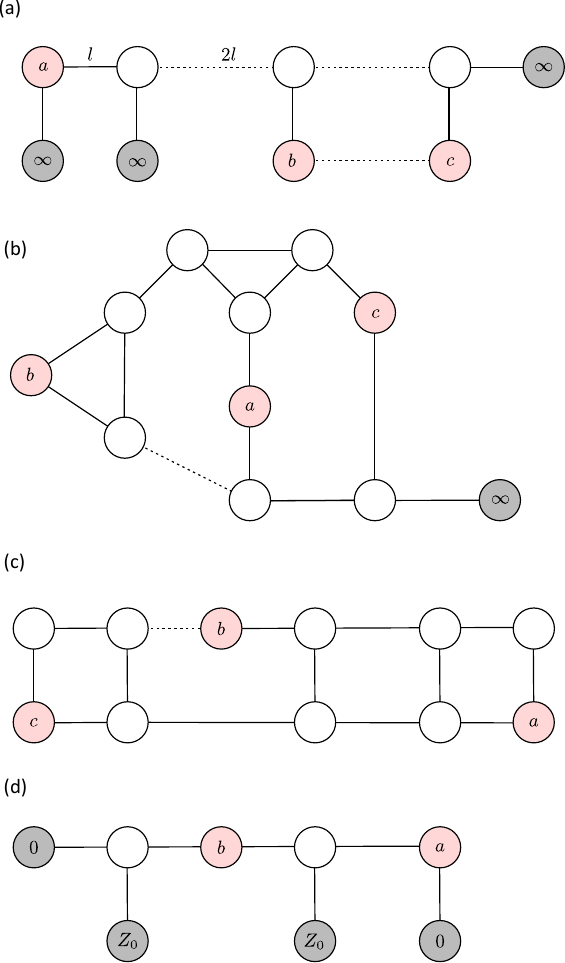}
\caption{\label{fig:Fig2}
Graph representation of different transmission line networks with non-geometric symmetries and thus non-trivial relations between scattering parameters when probing the network via the red-marked junctions $a,b,c$.
Lines represent coaxial cables of length $l$ or $2l$. 
Grey circles denote stubs, with termination resistance indicated ($0$ for short-circuit, $\infty$ for open-circuit, $Z_0$ for matched termination).
White and red circles denote junctions, with the important junctions $a,b,c$ marked in red.
} 
\end{figure} 

\PRLSec{More complicated examples}%
Non-geometric symmetries are not limited to two-port setups.
In \cref{fig:Fig2}(a) and (b), we show two networks of coaxial cables with $3$ ports.
Using the same techniques as before, for each network we can derive an effective $3\times{}3$ Hamiltonian $\mathcal{H}_s(k)$ that describes the dynamics at the red-highlighted junctions $s = \{a,b,c\}$.
The expressions for $\mathcal{H}_s(k)$ are cumbersome and shown in the Supplementary Material; what is important, however, is that both feature the same non-geometric symmetry $Q = \begin{psmallmatrix}
        0 & 1 & 1 \\
        1& 0 & 1 \\
        1& 1 & 0
\end{psmallmatrix}$; it follows that the scattering matrix of each setup fulfils
\begin{equation}
    \begin{aligned}
   r_b&=r_{c} -t_{a,b}+t_{a,c}\\
   r_a&=r_{c} -t_{a,b}+t_{b,c}
\end{aligned}
\end{equation}
\emph{for all frequencies}.

The relations can also have more complicated forms, as we show in \cref{fig:Fig2}(c); here, $Q = \begin{psmallmatrix}
  5 & 10 & -6 \\
 10 & -16 & 15 \\
 -6 & 15 & 0
\end{psmallmatrix}$, which leads to the relations
\begin{equation}
    \begin{aligned}
    r_b&=\frac{2}{5} t_{a,b}+\frac{2}{3} t_{a,c}-\frac{16}{15} t_{b,c}+r_{c}, \\
    r_a&=\frac{5}{2} t_{a,b}-\frac{5}{6} t_{a,c}-\frac{5}{3} t_{b,c}+r_{c}
\end{aligned}
\end{equation}

\PRLSec{Non-hermitian systems}%
Let us now investigate the effect of losses. So far, all cable stubs were terminated with either $Z=0$ or $Z=\infty$, neither of which introduces loss.
Moreover, for lossless cables, $k$ is real and the effective Hamiltonian is Hermitian.
When all cables are identical in type (though with possible different lengths) and lossy, $k$ becomes complex and the Hamiltonian non-Hermitian. However, this kind of uniform loss does not break the non-geometric symmetry (commutation with $Q$), so the broadband linear relation between scattering parameters remains intact.

A more interesting case is to consider non-uniform loss by terminating some stubs with resistors matching the cable impedance $Z_0$. An example is shown in \cref{fig:Fig2}(d), where two stubs are terminated by $Z_0$ resistors. The effective Hamiltonian of junctions $a,b$ commutes with
$ Q = 
\begin{psmallmatrix}
 0 & 1 \\
 1 & 1 \\
\end{psmallmatrix}$
and thus has a non-geometric symmetry. Consequently, the scattering parameters satisfy the broadband relation $r_a=r_b-t$.

\PRLSec{Generalization and applicability scope}%
From the above, the general principle is clear and surprisingly simple: if the effective Hamiltonian $\mathcal{H}_s(k)$ of a network has a (non-geometric) symmetry, the corresponding scattering matrix $S_s(k)$ shares this symmetry, resulting in linear relations between the scattering parameters that hold for all frequencies.
We remark that this principle has been discussed and used in Ref.~\cite{Sol2024AM362303891CovertScatteringControlMetamaterials}, though only for the special case of permutation matrices $Q$.

Although we have focused on coaxial cable networks, the above principle is applicable to a wide range of setups.
In the End matter, we demonstrate this for the two example cases of networks of acoustic pipes and RLC circuits.

Since it puts our results into a much broader context, let us explicitly derive them for quantum mechanical systems.
We consider a Hamiltonian $H$, describing a (possibly non-hermitian) system.
Coupling it to the environment, with coupling described by a matrix $W$, the scattering matrix reads \cite{sweeneyTheoryReflectionlessScattering2020}
\begin{equation}
    S(E) = I - 2 \pi i W^\dagger G(E) W \,,
\end{equation}
where
$G(E) :=  \left( E - H + i \pi W W^\dagger \right)^{-1}$ is the Green's function of the open system.
In the case where only a (small) part $s$ of the system is opened, this becomes
\begin{equation}
    S_s(E) = I - 2 \pi i W_s^\dagger G_{s}(E) W_s \,,
\end{equation}
with $G_{s} = \left(-\mathcal{H}_s(E) + i \pi W_s^\dagger W_s \right)^{-1}$, and
\begin{equation}
    \mathcal{H}_s = H_{ss} - EI - H_{s\overline{s}} \left( H_{\overline{ss}} - EI \right)^{-1} H_{\overline{s}s}
\end{equation}
the effective Hamiltonian of $H$ for the subsystem $s$, with $\overline{s}$ denoting its complement, that is, the remainder of the system.
Now, to make the connection to what we have done above, let us assume that the effective Hamiltonian has a (non-geometric) symmetry $\left[ Q,\mathcal{H}_s \right] = 0$.
If, additionally, the coupling is compatible with this symmetry, $Q W_s = W_s Q$, then $S_s$ commutes with $Q$, leading to broadband linear relations between scattering parameters.

We note that the effective Hamiltonian $\mathcal{H}_s$ is, up to an energy-shift, equal to the so-called isospectral reduction \cite{Bunimovich2014IsospectralTransformationsNewApproach}.
When $\mathcal{H}_s$ has a symmetry that $H$ itself does not feature (a mirror symmetry, for instance), it is said to feature a \emph{latent symmetry} \cite{Smith2019PA514855HiddenSymmetriesRealTheoretical}.
In the past years, such symmetries have been investigated in a range of physical setups, including topological Anderson insulators \cite{linTopologicalAndersonInsulators2026},  non-hermitian systems \cite{Cui2023PRL131237201ExperimentalRealizationStableExceptional,brandaoLatentSymmetryMinimal2026}, or waveguide networks \cite{Sol2024AM362303891CovertScatteringControlMetamaterials,Rontgen2023PRA20044082EquireflectionalityCustomizedUnbalancedCoherent,Rontgen2023PRL130077201HiddenSymmetriesAcousticWave}.

\begin{figure}[t]
    \centering
    \includegraphics[width=\linewidth]{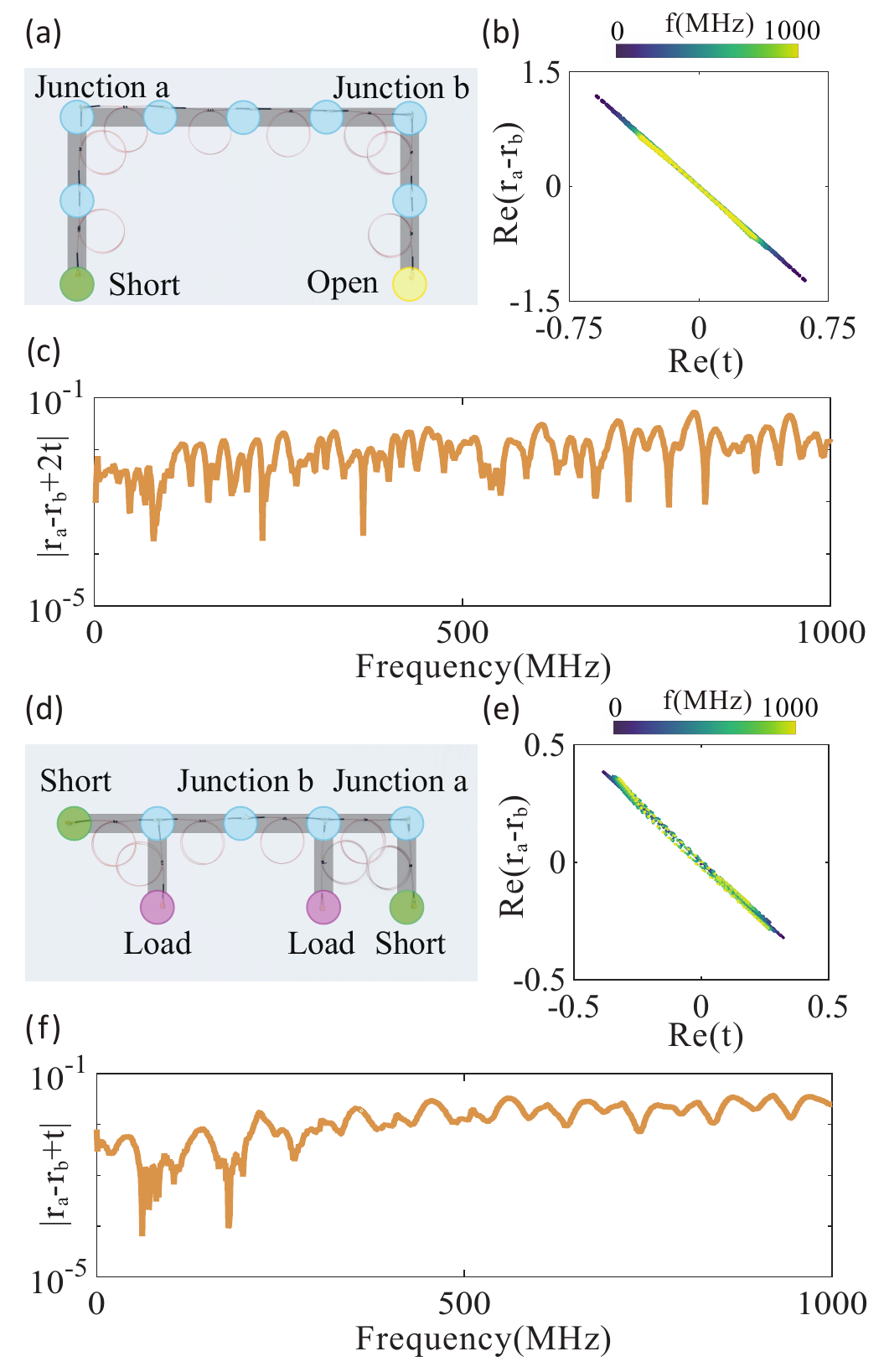}
    \caption{Experimental verification of the broadband linear scattering relation.
    (a) A photo of the experimental transmission line network for the model in
    Fig.~\ref{fig:Fig1}(a). (b) Experimentally measured relationship between
    $\operatorname{Re}(r_a-r_b)$ and $\operatorname{Re}(t)$. (c) Frequency dependence
    of the residual $|r_a-r_b+2t|$. (d)--(f) The same plots as
    in (a)--(c), but for the model in Fig.~\ref{fig:Fig2}(d).}
    \label{fig:Fig3}
\end{figure}

\PRLSec{Finding broadband symmetries}%
Let us now discuss how, given an effective Hamiltonian, its broadband symmetries (if any) can be found.
To this end, let us go back to our fundamental relation \cref{eq:effectiveCommutation}.
In practice, checking that this symmetry holds at \emph{every} frequency $k$ is made tractable by decomposing the effective Hamiltonian as
\begin{equation} \label{eq:hamEffDecomp}
    \mathcal{H}_s(k) = \sum_{i=1}^n A_i \, f_i(k) \,,
\end{equation}
where the $f_i(k)$ are linearly independent functions of $k$.
The $A_i$ are constant, $k$-independent matrices that describe how the particular $f_i(k)$ connects the ports $s$.

Because the $f_i$ are linearly independent, the condition $Q \mathcal{H}_s(k) = \mathcal{H}_s(k) Q$ holds for all $k$ if and only if $Q A_i = A_i Q$ for every $i$ individually.
Using the Sylvester equation, $Q$ is then obtained by finding the kernel of a small matrix; see End Matter for details.
To summarize, the task of finding a broadband symmetry of $\mathcal{H}_s$ -- and hence of $S_s$ -- is reduced to a finite, frequency-independent set of commutation relations, which can be checked directly on the matrices $A_i$ without any reference to $k$.

\PRLSec{Designing setups with broadband non-geometric symmetries}
Having treated the forward problem of determining the broadband symmetries of a given setup, let us now comment on the inverse problem.
That is, prescribing $Q$---and thus the desired relations in the scattering matrix---and designing a corresponding setup.
A suitable starting point is our decomposition \cref{eq:hamEffDecomp}.
With $Q$ chosen, the main idea is to determine allowed coefficient matrices $A_i$ and then build suitable cable, circuit or resonator motifs realizing these matrices; the final setup is then obtained by combining these motifs.

If one is not demanding a specific $Q$, but instead needs a large number of setups with any non-geometric symmetry, there is an alternative approach, explained in detail in the End Matter.
It is based on an exhaustive search on graphs, works surprisingly well and yielded us a large number of setups, such as those shown in \cref{fig:Fig2}.

\PRLSec{Experimental verification}%
To experimentally test our theoretical predictions, we realize the networks shown in Fig.~\ref{fig:Fig1}(a) and~\ref{fig:Fig2}(d) using coaxial-cable transmission-line networks (TLNs). TLNs provide a versatile platform for emulating wave networks and have been employed to realize topological phases \cite{jiangPhysicsTopologicalProperties2020,Jiang2019NC10434ExperimentalDemonstrationAngularMomentumdependent,jiangFourbandNonAbelianTopological2021,jiangObservationNonHermitianBoundary2024}, bound states in the continuum \cite{manzoorExperimentalObservationTimeDomain2026,wangBoundStatesContinuum2024}, and Möbius network geometries \cite{chenEigenmodeEigenpropagationElectromagnetic2024}.

For the experiments, we use 1-m-long SMA coaxial cables (BY-316-SMA-1). In Fig.~\ref{fig:Fig3}(a), short- and open-circuit boundary conditions are implemented using an SMA shorting cap and open termination, respectively. Links of length $l$ ($2l$) are formed by connecting two (four) cables in series, with small termination-induced differences in electrical length compensated through adjustment of the intermediate connectors. In Fig.~\ref{fig:Fig3}(d), each link of length $l$ is implemented by a single cable, while a $50\,\Omega$ SMA load realizes the load boundary condition.

We probe the two internal nodes $a$ and $b$ using a vector network analyzer (Ceyear 3671E), whose reference planes are calibrated to the two nodes. The complex scattering matrix $\boldsymbol{S}(f)$ is measured over $1\ \mathrm{MHz}$ to $1\ \mathrm{GHz}$. As shown in Fig.~\ref{fig:Fig3}(b), the measured $\mathrm{Re}(r_a - r_b)$ and $\mathrm{Re}(t)$ closely follow the predicted linear relation over the entire frequency range, and the corresponding residual remains close to zero [Fig.~\ref{fig:Fig3}(c)]. The load-terminated network exhibits the same behavior, as shown in Fig.~\ref{fig:Fig3}(e) and~\ref{fig:Fig3}(f).

\PRLSec{Concluding remarks}%
The frequency-independent scattering relations identified here point to a broader notion of symmetry in wave physics—one not tied to any geometric symmetry principle.
While demonstrated in waveguide networks, the underlying symmetries, and the linear relations they impose on scattering, are not specific to this platform: they offer a general tool for constraining and engineering scattering processes in classical and quantum wave systems alike. We anticipate that this framework will stimulate further theoretical work and guide the design of broadband devices built on these unconventional symmetries.

\section*{Acknowledgments}
This work was financially supported by the National Natural Science Foundation of China (NSFC), grant number W2533029, 12304345 and 12574345.
M.R. acknowledges financial support in the framework of the ``Eastern Institute for Advanced Study Postdoctoral Excellent Programme''.
M.R. acknowledges fruitful discussion with G.E. Sommer.
The authors are grateful to C.T. Chan for fruitful discussions as well as valuable feedback on the manuscript.

\section*{Author contributions}
M.R. conceived the project and performed most of the theoretical calculations, including the search for suitable networks.
X.C., S.L, C.Z., and R.Y. performed the measurements.
M.R. and S.L. wrote the manuscript.
T.J., W.G., and M.R. supervised the work.
All the authors discussed the results and contributed to this work.

\section*{Competing interests}
The authors declare no competing interests.


%

\clearpage
\appendix
\section{End Matter}

\PRLSec{Deriving the non-linear eigenvalue problem}%
We represent our network as a so-called quantum graph of $N$ vertices \cite{Berkolaiko2013186IntroductionQuantumGraphs,Kottos1999AoP27476PeriodicOrbitTheorySpectral,Kottos2003JPAMG363501QuantumGraphsSimpleModel}.
Each vertex is either a junction or a cable stub.
An edge between two vertices $n,m$ represents a cable of length $L_{nm}$, and acts as the domain for a one-dimensional Helmholtz equation \cref{eq:1DHelmholtz}.
Demanding Kirchhoff-like laws at junctions, as well as boundary-conditions at the stub, this allows to find the eigenmodes $V$ of the network.
We remark that quantum graphs can describe a variety of single-mode waveguides, such as supercooled microwave networks \cite{dietzClosedOpenSuperconducting2024}, coaxial cables \cite{cheExperimentalStudyDistributions2025,wangBoundStatesContinuum2024,lawniczakExperimentalNumericalInvestigation2008}, or networks of acoustic pipes \cite{lawrieApplicationQuantumGraph2024}, and have recently been used for the design of metamaterials \cite{lawrieQuantumGraphApproach2022,lawrieEngineeringMetamaterialInterface2023}.

Within each waveguide connecting vertices $n,m$, we can write the wave field 
\begin{equation} \label{eq:HelmholtzSolutionAnsatz}
    V_{nm}(x) = a e^{- i k x} + b e^{+ i k x}
\end{equation}
with $x$ the distance to $n$.
Using the $-i\omega t$ convention, $a$ ($b$) denotes the amplitude of the wave travelling towards (away from) vertex $n$.
Since the voltage field has to be continuous, we can assign a unique value $V_i$ to vertex $i$; this allows us to rewrite \cref{eq:HelmholtzSolutionAnsatz} as
\begin{equation} \label{eq:HelmholtzSolutionAnsatzRefined}
    V_{nm}(x) = \frac{V_n \sin\left( k(   L_{nm} - x)\right) + V_m \sin k x}{\sin k L_{nm}}
\end{equation}
with $x$ the distance to junction $n$ and $L_{nm}$ the length of the cable connecting junctions $n,m$.

At each junction $n$, the total current (proportional to the derivative $dV/dx$) has to vanish.
Thus, differentiating \cref{eq:HelmholtzSolutionAnsatzRefined} at $x=0$ and summing over all neighbours $m$ gives us
\begin{equation} \label{eq:currentSummation}
0 = \sum_m \frac{V_m - V_n \cos k L_{nm}}{\sin k L_{nm}} \,.
\end{equation}
So far, we did not take into account how the stubs are terminated; let us do that now.
Suppose one of the neighbors $m$ of vertex $n$ is a stub, terminated with $Z_m=0$; this corresponds to the Dirichlet boundary condition $V_m = 0$.
On the other hand, when $Z_m=Z_0$, the coefficient $a$ in \cref{eq:HelmholtzSolutionAnsatzRefined} has to vanish, translating into $V_m = exp(i k L_{nm}) V_n$.
Lastly, when $Z_m=\infty$ this corresponds to Neumann boundary conditions $V' = 0$.
Since $n$ is a stub, it has only one neighbour, and \cref{eq:currentSummation} then already encodes the Neumann boundary condition.

After incorporating this into \cref{eq:currentSummation}, we see that all $V_m$ with $Z_m = 0, Z_0$ are eliminated.
With a sharp eye, we see that \cref{eq:currentSummation} is nothing but the $V_n$ entry of the non-linear eigenvalue problem 
\begin{equation} \label{eq:fullNEVP}
    \mathcal{H} \vec{V} = 0 \,.
\end{equation}
Here, $\mathcal{H}$ is a matrix of dimension $N-n_{0} - n_M$, where $n_0$, $n_M$ denote the number of stubs terminated with $Z= 0$ and $Z=Z_0$ (matched end), respectively.
The entries of $\mathcal{H}$ are thus
\begin{equation}
    \left(\mathcal{H}\right)_{nm} = \begin{cases}
h_{nn} & n=m \\
g_{nm} & \, n\ne m
\end{cases}
\end{equation}
with $g_{nm}=1/\sin k L_{nm}$ if $n,m$ are neighbours and $g_{nm}=0$ otherwise, and with
\begin{equation}
    h_{nn} = -\sum_{m'} \cot k L_{nm'} + \sum_{m': Z_0} \left(i + \cot k L_{nm'} \right)
\end{equation}
where the first sum goes over all neighbours of vertex $n$ in the network graph, and the second only over those terminated by $Z_0$.

Lastly, choosing a relevant vertex set $s$, the effective Hamiltonian $\mathcal{H}_s$ can be obtained from \cref{eq:fullNEVP} by eliminating all $V_j$ with $j \notin s$; that is, the complement $V_{\overline{s}}$.
This can be achieved by first reordering the rows and columns of  \cref{eq:fullNEVP} such that it has the block-form
\begin{equation} \label{eq:blockNEVP}
\begin{pmatrix}
        \mathcal{H}_{ss} & \mathcal{H}_{s\overline{s}} \\
        \mathcal{H}_{\overline{s}s} & \mathcal{H}_{\overline{ss}}
\end{pmatrix} \cdot
\begin{pmatrix}
    V_s \\
    V_{\overline{s}}
\end{pmatrix} = 0 \,.
\end{equation}
Here, $\mathcal{H}_{ab}$ denotes the matrix obtained from $\mathcal{H}$ by taking the rows in the set $a$, and the columns in the set $b$.
Multiplying out \cref{eq:blockNEVP} yields two coupled equations; taking the second, solving for $V_{\overline{s}}$ and inserting into the first gives
\begin{equation}
  \mathcal{H}_s(k) V_s = 0  
\end{equation}
with the effective Hamiltonian being the Schur complement
\begin{equation} \label{eq:SchurComplement}
    \mathcal{H}_s = \mathcal{H}_{ss} - \mathcal{H}_{s\overline{s}} \left( \mathcal{H}_{\overline{ss}}\right)^{-1} \mathcal{H}_{\overline{s}s} \,.
\end{equation}

\PRLSec{Theory for acoustic waveguide networks}%
Networks of thin, identical acoustic waveguides are governed by the one-dimensional Helmholtz equation \cref{eq:1DHelmholtz}, with $V$ replaced by the pressure field $p$ \cite{Coutant2021PRB103224309AcousticSuSchriefferHeegerLatticeDirect,Rontgen2023PRL130077201HiddenSymmetriesAcousticWave}.
At each junction, the pressure field $p$ must be continuous, and the total flux (proportional to $p'$) must vanish.
Thus, \cref{eq:currentSummation} holds when replacing $V_j$ by $p_j$, the pressure at junction $j$.

As for the termination of stubs, the situation is more complicated.
Firstly, we remind the reader that termination of a coaxial cable with open end $Z = \infty$ leads to Neumann boundary conditions, $V'=0$.
This has a convenient partner in acoustic waveguide networks, where Neumann boundary conditions are naturally imposed at the boundaries of a closed waveguide, as the flux has to vanish there.
Our two other terminations, $Z=0$, $Z=Z_0$ are more difficult to achieve in acoustics; one would need to achieve Dirichlet ($p=0$) or matched boundary conditions in a broad frequency band to obtain acoustic analogues.
We leave these problems to specialists and note that, even when taking only networks with Neumann boundary conditions, there are still many that can be implemented in acoustics; a particularly simple one would be the networks of \cref{fig:Fig2}(a),(b), and (c).

\PRLSec{Theory for RLC circuits}%
In the past few years, circuit networks consisting of resistors, inductances, and capacitors became more and more popular in physics as a versatile platform for realising topological phenomena; see for instance Ref.˜\cite{Lee2018CP11TopolectricalCircuits,Imhof2018NP14925TopolectricalcircuitRealizationTopologicalCorner,jiangEngineeringTopolectricalCircuits2026}.
Using Kirchhoff's laws and Ohm's law, and working in frequency domain, the eigenmodes of such a circuit obey the non-linear eigenvalue problem
\begin{equation}
    \left(R + \frac{L}{i \omega} + i \omega C\right) \vec{V} = 0
\end{equation}
with the vector $\vec{V}$ denoting the voltage at nodes, and with the matrices $R,L,C$ describing how the nodes are interconnected in terms of resistors ($R$), inductances ($L$), and capacitors ($C$).
Completely analogous to \cref{eq:blockNEVP}, we can create an effective model for some nodes $s$ of relevance by employing the Schur complement, \cref{eq:SchurComplement}, on
\begin{equation}
    \mathcal{H} = R + \frac{L}{i \omega} + i \omega C\,.
\end{equation}
The circuit scattering matrix can then be computed as
\begin{equation}
    S_s = (I + \mathcal{H}_s^{-1})^{-1} (I - \mathcal{H}_s^{-1}) 
\end{equation}
and we see that, again, broadband non-geometric symmetries of $\mathcal{H}_s$ induce broadband linear relations between the matrix elements of $S_s$.

\PRLSec{Finding broadband symmetries using the Sylvester equation}
Any matrix $Q$ that simultaneously commutes with all the $A_i$ from \cref{eq:hamEffDecomp} needs to fulfil the Sylvester equation
\begin{equation} \label{eq:sylvesterEquation}
    \left(I \otimes X + Y \otimes I\right) \text{vec}(Q) = 0 \,,
\end{equation}
where $\text{vec}$ is the vectorization operator, and with the matrices $X = \left(A_{(1)}^T,\ldots,A_{(n)}^T\right)^T$, $Y =  \left(A_{(1)}^T,\ldots,A_{(n)}^T\right)$.
Finding $Q$ then amounts to finding the kernel of the small matrix $\left(I \otimes X + Y\otimes I\right)$, the dimension of which is just $(n |s|^2) \times (|s|^2)$, with $|s|$ size of the set $s$.

\PRLSec{Exhaustive search for finding systems with non-geometric latent symmetries}
We start with a simple graph, equate all edge lengths, and compute its effective Hamiltonian $\mathcal{H}_s(k)$ for some subset of vertices $s$; this task can be trivially parallelized.
To find suitable candidates and spare ourselves some work, we then evaluate, for each set $s$, $\mathcal{H}_s(k)$ at a few different frequencies $k_1,\ldots, k_n$, yielding matrices $\mathcal{H}^{(1)},\ldots{}, \mathcal{H}^{(n)}$.
Using the Sylvester equation approach from the End Matter, we then find the list of matrices that simultaneously commute with all of the $\mathcal{H}^{(i)}$.
Filtering out all permutation matrices, we are left with non-geometric broadband symmetries.

To feed this machinery, we need to generate large numbers of graphs; this can be very efficiently done using established tools from graph theory; in particular, using the nauty suite \cite{McKay2014JoSC6094PracticalGraphIsomorphismII}.
If a graph $G$ contains stubs---vertices connected to only one other vertex---, we need to decide the termination impedance condition ($0$, $\infty$, or $Z_0$); a graph with $n$ stubs thus splits into $3^n$ different configurations that need to be tested individually.

\clearpage
\onecolumngrid
\section{Supplemental Material}

\subsection{Effective Hamiltonians for networks of \cref{fig:Fig2}}
In the following, $c_n = \cot(n k)$, and $s_n = \sin(n k)$; we have also set $l=1$.

The effective Hamiltonians of subfigures (a) and (b) both have the form
\begin{equation}
    \mathcal{H}_s = \left(
\begin{array}{ccc}
 h_{1,1} & h_{1,2} & h_{1,3} \\
 h_{1,2} & h_{2,2} & h_{1,1}+h_{1,3}-h_{2,2} \\
 h_{1,3} & h_{1,1}+h_{1,3}-h_{2,2} & h_{1,2}-h_{1,3}+h_{2,2} \\
\end{array}
\right) \,.
\end{equation}

For subfigure (a), the expressions are
{\footnotesize
\begin{align*}
     h_{1,1} &= \frac{(-5248 c_1+2264 c_2-4310 c_3+1584 c_4-3114 c_5+648 c_6-1863 c_7-729 c_9+1136) \csc (k)}{5174 c_2+3816 c_4+2106 c_6+729 c_8+2831}\\
     h_{1,2} &= \frac{4 (310 c_1+299 c_2+3 (74 c_3+72 c_4+36 c_5+27 c_6+60)) \csc (k)}{5174 c_2+3816 c_4+2106 c_6+729 c_8+2831}\\
     h_{2,2} &= -\frac{(-7024 c_1+25750 c_2-5488 c_3+21192 c_4-3600 c_5+14175 c_6-1296 c_7+7290 c_8+2187 c_{10}+13502) \csc (k) \sec (k)}{4 (5174 c_2+3816 c_4+2106 c_6+729 c_8+2831)}\\
     h_{1,3} &= \frac{4 (340 c_1+116 c_2+237 c_3+90 c_4+135 c_5+74) \csc (k)}{5174 c_2+3816 c_4+2106 c_6+729 c_8+2831}
\end{align*}
}

For subfigure (b), the expressions are
{\footnotesize
\begin{align*}
     h_{1,1} &= -\frac{(16054 c_1-4336 c_2+13524 c_3-2304 c_4+10044 c_5-1296 c_6+5103 c_7+2187 c_9-2048) \cot (k)}{2 (3 c_1+1) (-2 c_1+3 c_2+1) (88 c_1+957 c_2+72 c_3+594 c_4+243
   c_6+574)}\\
     h_{1,2} &= \frac{2 (394 c_1+32 c_2+255 c_3+36 c_4+135 c_5+28) \csc (k)}{(-2 c_1+3 c_2+1) (88 c_1+957 c_2+72 c_3+594 c_4+243 c_6+574)}\\
     h_{2,2} &= -\frac{(8190 c_1-6056 c_2+6688 c_3-3936 c_4+4320 c_5-1944 c_6+2025 c_7-486 c_8+729 c_9-3386) \csc (k)}{2 (-2 c_1+3 c_2+1) (88 c_1+957 c_2+72 c_3+594 c_4+243
   c_6+574)}\\
     h_{1,3} &= \frac{2 (2411 c_1+159 c_2+1746 c_3+126 c_4+864 c_5+81 c_6+243 c_7+82) \csc (k)}{(3 c_1+1) (-2 c_1+3 c_2+1) (88 c_1+957 c_2+72 c_3+594 c_4+243 c_6+574)}
\end{align*}
}

For subfigure (c), the effective Hamiltonian is given by
{\footnotesize
\begin{equation}
    \mathcal{H}_s = \left(
\begin{array}{ccc}
 \frac{-16 \tan (k)-272 \cot (k)+462 s_{2}+270 s_{4}+162 s_{6}}{207 c_2+108 c_4+81 c_6+100} & \frac{8 (7 (\tan (k)-9 s_{2})+67 \cot (k))}{621 c_2+324 c_4+243 c_6+300} &
   \frac{280 \cot (k)-8 (\tan (k)+18 s_{2})}{621 c_2+324 c_4+243 c_6+300} \\
 \frac{8 (7 (\tan (k)-9 s_{2})+67 \cot (k))}{621 c_2+324 c_4+243 c_6+300} & \frac{-44 \tan (k)-384 \cot (k)+606 s_{2}+270 s_{4}+162 s_{6}}{207 c_2+108 c_4+81 c_6+100} &
   \frac{76 (\tan (k)-9 s_{2})+616 \cot (k)}{621 c_2+324 c_4+243 c_6+300} \\
 \frac{280 \cot (k)-8 (\tan (k)+18 s_{2})}{621 c_2+324 c_4+243 c_6+300} & \frac{76 (\tan (k)-9 s_{2})+616 \cot (k)}{621 c_2+324 c_4+243 c_6+300} & \frac{-68 \tan (k)-896 \cot
   (k)+1386 s_{2}+810 s_{4}+486 s_{6}}{621 c_2+324 c_4+243 c_6+300} \\
\end{array}
\right)
\end{equation}
}

Lastly, for subfigure (d), the effective Hamiltonian is given by
\begin{equation}
    \mathcal{H}_s = \left(
\begin{array}{cc}
 -2 \cot (k)+\frac{\csc ^2(k)}{2 \cot (k)-i} & \frac{\csc ^2(k)}{2 \cot (k)-i} \\
 \frac{\csc ^2(k)}{2 \cot (k)-i} & \frac{2-2 \cot (k) (\cot (k)-i)}{2 \cot (k)-i} \\
\end{array}
\right) \,.
\end{equation}

\subsection{Experimental implementation of the transmission-line networks}

The experimental networks are implemented using $50\Omega$ SMA coaxial cables and SMA T-junctions. Within the frequency range considered here, the cables operate in the fundamental transverse electromagnetic (TEM) mode and are well described by the standard transmission-line model \cite{Pozar2012MicrowaveEngineering,collin2007foundations}. Propagation through a cable of length $l$ introduces the factor

\begin{equation}
e^{-\gamma l}=e^{-\alpha l}e^{-j\beta l},
\end{equation}

where $\alpha$ and $\beta$ are the attenuation and phase constants, respectively. Thus, the physical length of each cable section determines the propagation phase accumulated along the corresponding network link. Losses and weak impedance discontinuities associated with cables, connectors, and junctions are naturally included in the experimentally measured scattering response\cite{dunsmore2020handbook}.

The boundary conditions used in our experiments are realized by terminating the corresponding transmission-line branches with open-circuit, short-circuit, or $50 \Omega$ matched loads. For a termination with impedance $Z_L$, the reflection coefficient is 

\begin{equation}
\Gamma_L=\frac{Z_L-Z_0}{Z_L+Z_0},
\end{equation}

such that

\begin{equation}
\Gamma_{\mathrm{open}}=1,
\qquad
\Gamma_{\mathrm{short}}=-1,
\qquad
\Gamma_{\mathrm{load}}\simeq0.
\end{equation}

These terminations therefore provide the experimental counterparts of the boundary conditions imposed in the theoretical network model.

In practice, small differences in the length of nominally equivalent paths arise from cable tolerances, SMA connectors, adapters, and terminations. For a TEM transmission line, electrical length mismatch $\Delta l$ produces an additional phase shift

\begin{equation}
\Delta\phi=\beta\Delta l = \frac{2\pi f\sqrt{\epsilon_r}}{c}\Delta l.
\end{equation}

These residual phase differences could be compensated using a variable RF phase shifter inserted into the corresponding transmission path. The phase shift is adjusted to restore the designed relative propagation phase between the network links, thereby minimizing deviations caused by uncontrolled electrical-length differences\cite{Pozar2012MicrowaveEngineering,dunsmore2020handbook}. 

In our experiments, we performed the compensation by slightly rotating the screws on the connectors, thereby inducing slight length changes.

\end{document}